\documentclass[%
 reprint,
 amsmath,amssymb,
 aps,prd]{revtex4-2}

\usepackage{graphicx}
\usepackage{dcolumn}
\usepackage{bm}

\begin{document}

\preprint{APS/123-QED}

\title{Unique and Universal Effects of Oscillation in Eccentric Orbital Binary Black Hole Mergers beyond Orbital Averaging}

\author{Hao Wang}
\email{husthaowang@hust.edu.cn}
\affiliation{Department of Astronomy, School of Physics, Huazhong University of Science and Technology, Wuhan 430074, China}

\author{Yuan-Chuan Zou}
\email{zouyc@hust.edu.cn}
\affiliation{Department of Astronomy, School of Physics, Huazhong University of Science and Technology, Wuhan 430074, China}

\author{Qing Wen Wu}
\email{qwwu@hust.edu.cn}
\affiliation{Department of Astronomy, School of Physics, Huazhong University of Science and Technology, Wuhan 430074, China}

\date{\today}

\begin{abstract}
We analyze 192 sets of binary black hole merger data in eccentric orbits obtained from RIT, decomposing the radiation energy into three distinct phases through time: inspiral, late inspiral to merger, and ringdown. Our investigation reveals a universal oscillatory behavior in radiation energy across these phases, influenced by varying initial eccentricities.
From a post-Newtonian perspective, we compare the orbital average of radiation energy with the non-orbital average during the inspiral phase. Our findings indicate that the oscillatory patterns arise from non-orbital average effects, which disappear when orbital averaging is applied. This orbital effect significantly impacts the mass, spin, and recoil velocity of the merger remnant, with its influence increasing as the initial eccentricity rises. Specifically, in the post-Newtonian framework, the amplitudes of oscillations for mass, spin, and recoil velocity at ${e_t}_0 = 0.5$ (initial temporal eccentricity of PN) are enhanced by approximately 10, 5, and 7 times, respectively, compared to those at ${e_t}_0 = 0.1$. For a circular orbit, where ${e_t}_0 = 0.0$, the oscillations vanish entirely.
These findings have important implications for waveform modeling, numerical relativity simulations, and the characterization of binary black hole formation channels.
\end{abstract}

\maketitle

\noindent {\textbf{\textit{Introduction}}.}
The circularization effect of gravitational radiation is the primary reason behind the use of circular orbit templates by the LVK (LIGO \cite{LIGOScientific:2014qfs}, Virgo \cite{VIRGO:2014yos}, KAGRA \cite{KAGRA:2018plz}) gravitational wave detectors for detection \cite{Peters:1963ux,Peters:1964zz}. However, in densely populated stellar environments, such as globular clusters \cite{Miller:2002pg,Rodriguez:2016kxx} or galactic nuclei \cite{Gondan:2020svr}, certain dynamical formation channels can lead to the emergence of binary star systems with eccentric orbits \cite{Gultekin:2005fd,Samsing:2013kua,Samsing:2017oij}. Among the observed events, GW190521 \cite{LIGOScientific:2020iuh} stands out as a potential example of a dynamical binary black hole merger detected to date \cite{Gayathri:2020coq}.

Post-Newtonian (PN) theory, a pivotal component of the analytic relativity, has progressively refined the modeling of the dynamics and waveforms of binary black holes (BBH) in eccentric orbits with increasing precision over the past decade \cite{Blanchet:2013haa}. However, its efficacy diminishes in the vicinity of the merger moment, necessitating recourse to numerical relativity (NR).
Research in NR concerning BBH in eccentric orbits has witnessed significant advancements in the last decade. These strides include the analysis of the transition from inspiral to plunge in eccentric orbits \cite{Sperhake:2007gu}, investigations into orbital circularization \cite{Hinder:2007qu}, examinations of the recoil, mass, and spin of remnants in low eccentricity orbits \cite{Huerta:2019oxn}, explorations of kick enhancement attributed to eccentricity \cite{Sperhake:2019wwo}, studies on anomalies in recoil due to eccentricity \cite{Radia:2021hjs}, and the extension to eccentric orbits in RIT's fourth data release \cite{Healy:2022wdn}. Collectively, these investigations are gradually unveiling the intricate orbital effects of eccentricity.

In our previous work \cite{Wang:2023vka}, we provided a comprehensive summary and analysis of the dynamic properties of eccentric orbit BBH mergers from RIT \cite{Healy:2022wdn}. This analysis included quantities such as peak luminosity, the mass, spin, and recoil velocity of the remnant. We observed that when the initial eccentricity in the NR simulations was sufficiently dense or the NR simulations with continuously changing initial eccentricity are sufficient, these dynamic quantities exhibited oscillatory behavior with changing initial eccentricity. By associating the integer value of the orbital cycle with the peaks and valleys of this oscillation, we made significant progress explain this phenomenon. However, given the oscillatory correlation between the peak luminosity and eccentricity, coupled with the predominance of radiation during the merger phase, we attribute this phenomenon to the profound influence of strong field effects in the merger phase.

In this paper, we aim to conduct a more detailed re-examination of the mechanism behind the generation of these oscillations, delving deeper into the perspectives of PN and NR. Throughout this paper, we adopt geometric units where $G=c=1$. The component masses of BBH are represented as $m_1$ and $m_2$, while the total mass is denoted by $M$. We set the total mass $M$ at unity (although occasionally we explicitly write it for clarity). The mass ratio $q$ is defined as $q = m_1/m_2$, with $m_1 \leq m_2$.

\begin{figure}[htbp!]
\centering
\includegraphics[width=8cm,height=5cm]{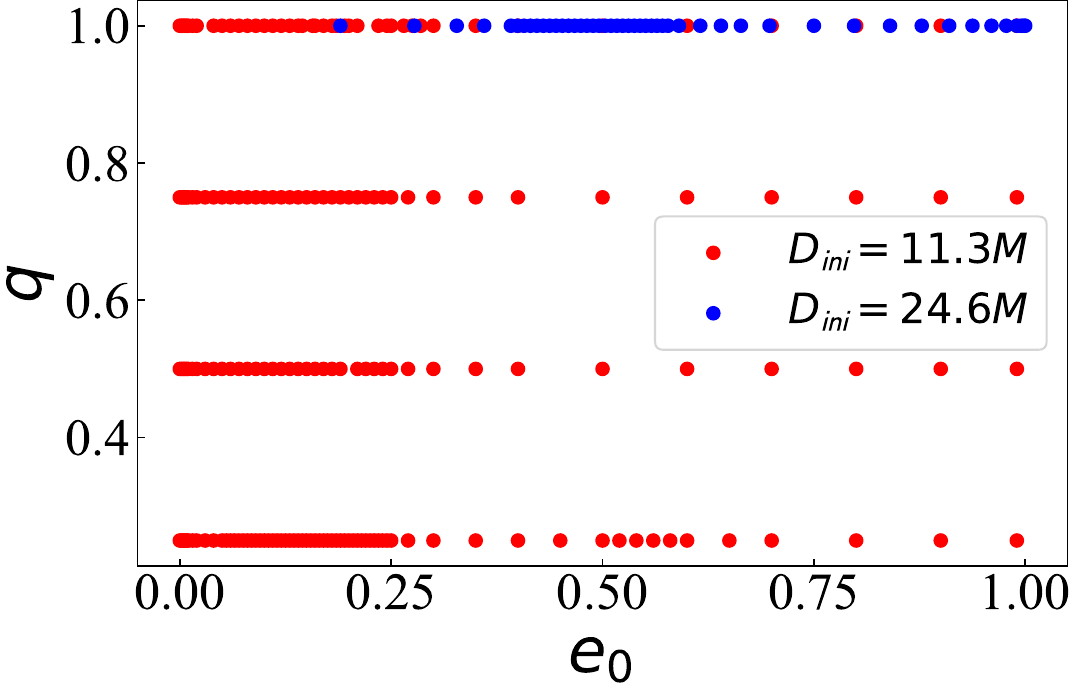}
\caption{\label{FIG:1}Parameter space of the eccentric nonspinning BBH NR simulations used in our study. There are 192 sets simulations with $D_{\text{ini}}=11.3M$ and $q=1$ (43 sets), $q=0.25$ (67 sets), $q=0.5$ (41 sets), $q=0.75$ (41 sets), and 48 sets simulations with $D_{\text{ini}}=24.6M$ and $q=1$.}
\end {figure}

\noindent {\textbf{\textit{Effects of Oscillation}}.}
We meticulously selected 192 sets of RIT waveforms showcasing oscillations, all nonspinning, with fixed initial coordinate distances $D_\text{ini}$ of $11.3M$ and $24.6M$, accompanied by mass ratios of $q=1, 0.75,0.5,0.25$. Within FIG. \ref{FIG:1}, we delineate the parameter space of the eccentric nonspinning BBH NR simulations used in our study.
The initial eccentricity $e_0$ is derived from RIT's eccentricity measurement methodology \cite{Healy:2022wdn}. Notably, the eccentricities of waveforms at $D_\text{ini} = 11.3M$ predominantly reside in the low eccentricity range (0-0.25), while those at $D_\text{ini} = 24.6M$ cluster within the medium eccentricity range (0.3-0.6), a prerequisite for uncovering the oscillation effect \cite{Wang:2023vka}. In the interest of brevity, we specifically focus on the remnant's mass $M_\text{rem}$, omitting presentations on the spin $\alpha_\text{rem}$, and recoil velocity $V_\text{rem}$ because these dynamic quantities stem from either the Newman-Penrose scalar $\Psi_4$ or gravitational wave strain $h$, with similar formulas and origins. The remnant's mass is computed as $M_\text{rem}=M_{\mathrm{ADM}}-E_{\mathrm{rad}}$, where $M_{\mathrm{ADM}}$ denotes the ADM (Arnowitt, Deser, Misner) mass and $E_{\mathrm{rad}}$ signifies the gravitational radiation energy.

\begin{figure}[htbp!]
\centering
\includegraphics[width=8cm,height=5cm]{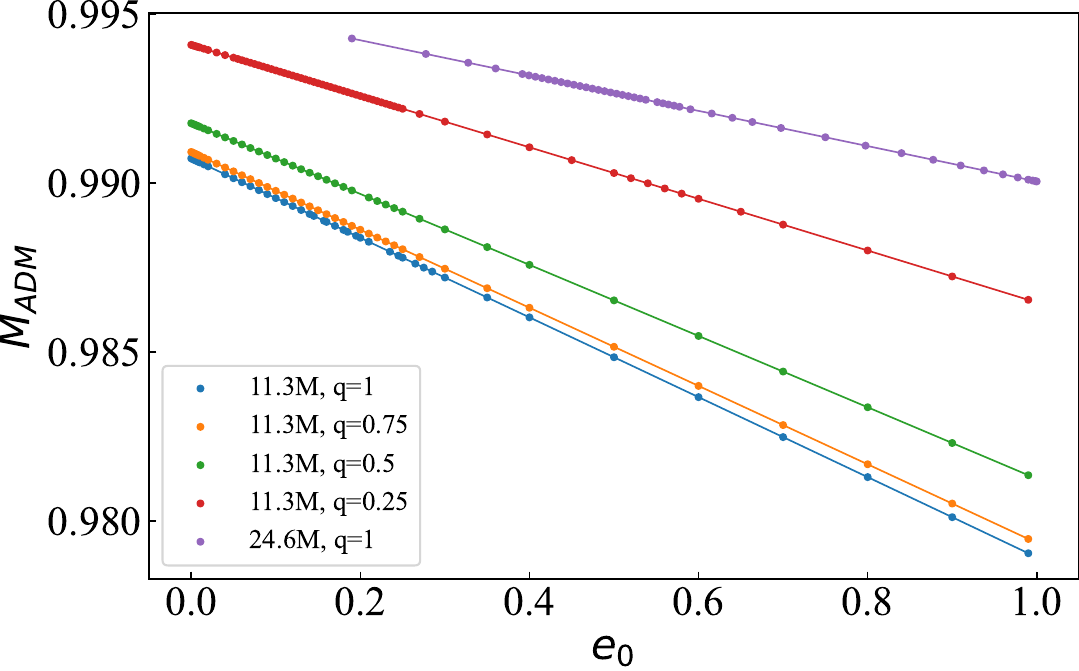}
\caption{\label{FIG:2}Relationship between initial ADM mass $M_{\mathrm{ADM}}$ and initial eccentricity $e_0$ for the parameter space in the FIG. \ref{FIG:1}.}
\end {figure}

RIT initializes the eccentric orbit BBH merger data by fixing the initial distance $D_\text{ini}$ and manipulating the tangential linear momentum $p_\text{t}$. In contrast, Ref. \cite{Radia:2021hjs} achieves this by fixing the binding energy $E_{\mathrm{b}}$, obtained through $E_{\mathrm{b}}=M_{\mathrm{ADM}}-M$, and adjusting $p_\text{t}$. Despite differing methodologies, both approaches produce the same oscillation effect due to the continuous alteration of initial eccentricity.
FIG. \ref{FIG:2} illustrates the relationship between initial ADM mass and initial eccentricity for the parameter space in the FIG. \ref{FIG:1}, indicating a gradual change without oscillation. Thus, the oscillation phenomenon arises from the radiated energy $E_{\mathrm{rad}}$ rather than ADM mass fluctuations. For angular momentum $L$, similar to energy, the oscillation does not come from the initial angular momentum $L_0$. We will not go into too much detail here. In NR, $E_{\mathrm{rad}}$ can be evaluated through \cite{Ruiz:2007yx}
\begin{equation}\label{eq:1}
E_{\mathrm{rad}}(t)=\lim _{r \rightarrow \infty} \frac{r^2}{16 \pi}\sum_{\ell, m} \int_{t_0}^t \mathrm{~d} t^{\prime}\left|\dot{h}^{\ell m}\right|^2,
\end{equation}
where $h^{\ell m}$ represents the harmonic mode post-decomposition of $h$ by the spin-weighted spherical harmonic function, $\dot{h}^{\ell m}$ denotes the time derivative of $h^{\ell m}$, and $r$ signifies the radius for waveform extraction. Through time interval control in the energy computation process, we categorize the outcomes into three distinct phases: the inspiral phase spanning from the initial moment $t_0$ to $-200M$ (with $200M$ preceding the merger, where we designate the merger time as 0); the late inspiral to the merger, covering $-200M$ to 0; and the subsequent ringdown phase, commencing beyond 0.  

\begin{figure*}[htbp!]
\centering
\includegraphics[width=16cm,height=10cm]{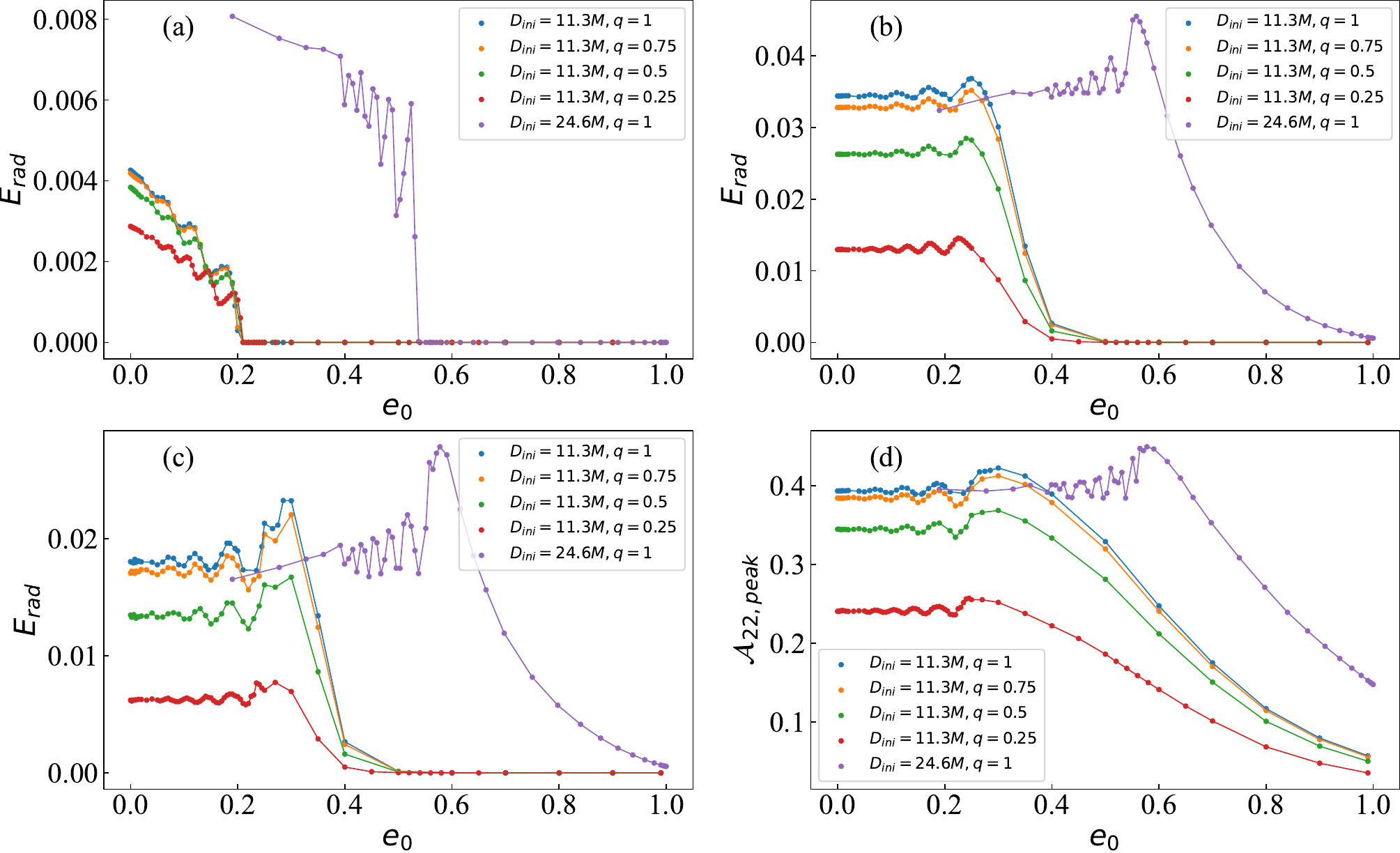}
\caption{\label{FIG:3}Correlation between the computed radiation energy $E_{\mathrm{rad}}$ and the initial eccentricity $e_0$ across three time  intervals: inspiral phase spanning from the initial moment $t_0$ to $-200M$ (some waveforms have a radiation of 0 because the initial time is greater than $-200M$) (panel (a)); the late inspiral to the merger, covering $-200M$ to 0 (panel (b)); and the subsequent ringdown phase, commencing beyond 0 (panel (c)), along with the association between the peak amplitude $\mathcal{A}_{22,\text{peak}}$ and $e_0$ in panel (d).}
\end {figure*}

In FIG. \ref{FIG:3}, we present the correlation between the computed radiation energy across three intervals and the initial eccentricity in panels (a), (b) and (c), along with the association between the amplitude of the 22 mode at the merging instant or the peak amplitude $\mathcal{A}_{22,\text{peak}}$ and $e_0$ in panel (d). It is important to emphasize that the time reference does not have to be restricted to $-200M$; similar results are observed at $-300M$, where oscillations still persist. The 22 mode amplitude $\mathcal{A}_{22}$ is derived by decomposing $h^{\ell m}$ into amplitude and phase components, $h^{\ell m}=\mathcal{A}^{\ell m}(t) \exp \left[-i \Phi^{\ell m}(t)\right]$.

Panel (a) reveals oscillations during the inspiral phase, albeit with an overall downward trend, attributed to the diminishing $M_\text{ADM}$ as depicted in FIG. \ref{FIG:2}. The oscillations in panels (b), (c), and (d) exhibit minimal influence from $M_\text{ADM}$, showcasing consistent horizontal oscillations akin to the remnant mass $M_\text{rem}$ as observed in Ref. \cite{Wang:2023vka}. Aligning the values in panel (a) with the other panels in FIG. \ref{FIG:3}, we discern that the oscillations in the inspiral phase mirror the patterns in the other panels, featuring analogous peaks and valleys. All of results in FIG. \ref{FIG:3} implies that the impact of initial eccentricity extends through the inspiral phase, the late inspiral to merger phase, moment of merger (where the radiation energy computation hinges on the amplitude), and the ensuing ringdown phase.
Consequently, altering the eccentricity not only influences the weak field dynamics during the inspiral phase of BBH mergers but also affects the strong field dynamics at the merger phase. This influence engenders a consistent oscillatory effect, indicating a unified origin for the impact of eccentricity on BBH mergers.

In our prior study \cite{Wang:2023vka}, we linked the peaks and valleys of the oscillations to integer gravitational wave orbital cycle numbers $N_{\text {orbits }}=\Delta \Phi^{22} /(4\pi)$, where $\Delta \Phi^{22}=\Phi^{22}\left(t_{\mathrm{merger}}\right)-\Phi^{22}\left(t_0+t_{\mathrm{relax}}\right)$ (with $t_{\mathrm{merger}}$ denoting the merger time and $t_{\mathrm{relax}}$ as the relaxation time of junk radiation). This analysis highlighted that the oscillations stem from the influence of integer orbits, with the peaks and valleys aligning with orbital transitions. While this insight suggested a connection to the orbital dynamics, it did not provide a comprehensive explanation for their origin.

Nevertheless, the existence of the same oscillations during the inspiral phase in panel (a) of FIG. \ref{FIG:3} offers us an opportunity to delve into this phenomenon from a PN perspective.

\noindent {\textbf{\textit{Post-Newtonian comparison}}.}
In our previous work \cite{Wang:2024jro}, we present a comprehensive comparison between the eccentric orbital waveforms generated by RIT \cite{Healy:2017psd,Healy:2019jyf,Healy:2020vre,Healy:2022wdn} and SXS \cite{Mroue:2013xna,Boyle:2019kee} and those derived from PN calculations. The PN waveform model utilized in our analysis is grounded on a 3PN order quasi-Kepler parameterization for conservative dynamics \cite{Hinder:2008kv,Memmesheimer:2004cv}, incorporating the radiative reaction of 3PN that includes both instantaneous and hereditary terms \cite{Arun:2009mc,Arun:2007rg,Arun:2007sg}, and extending to the gravitational wave amplitude of 3PN beyond the quadrupole moment \cite{Mishra:2015bqa,Boetzel:2019nfw,Ebersold:2019kdc}.

Through a meticulous least squares fitting approach applied to the frequencies of the 22 mode of $\Psi_4$ at $-200M$ prior to the merger (selected due to PN model limitations beyond this point, with the frequency around 0.1), we derive the initial PN parameters $x_0$, ${e_t}_0$, and $l_0$ \cite{Wang:2024jro}. Here, $x$ represents the PN expansion parameter, $e_t$ signifies temporal eccentricity, and $l$ denotes the mean anomaly.

\begin{figure}[htbp!]
\centering
\includegraphics[width=8cm,height=5cm]{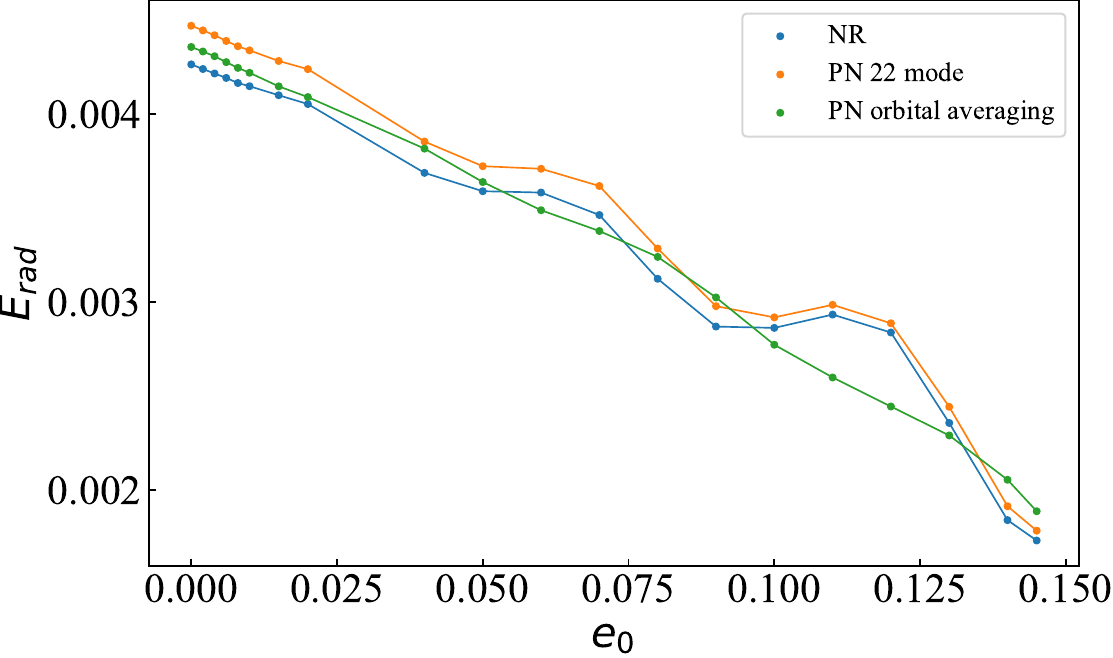}
\caption{\label{FIG:4}Relationship between initial eccentricity $e_0$ and NR radiation energy (corresponding to panel (a) of $q=1$ and $D_\text{ini}=11.3M$ in FIG. \ref{FIG:3}), the radiation energy computed through the orbital average energy flux of PN, and the PN quadrupole radiation energy of the gravitational wave mode 22 derived using Eq. (\ref{eq:1}).}
\end {figure}

In FIG. \ref{FIG:4}, we conduct a comparison between NR and PN radiation energy for mass ratio of $q=1$ and $D_\text{ini}=11.3M$ as a representative case. The methodology and outcomes for different mass ratios and the scenario involving radiation angular momentum follow a similar approach. Within FIG. \ref{FIG:4}, we illustrate the relationship between initial eccentricity $e_0$ and NR radiation energy (corresponding to panel (a) in FIG. \ref{FIG:3}), the radiation energy computed through the orbital average energy flux of PN \cite{Arun:2007sg}, and the PN quadrupole radiation energy of the gravitational wave mode 22 derived using Eq. (\ref{eq:1}).
The focus on the quadrupole moment's 22 mode stems from findings in Ref. \cite{Wang:2024jro}, indicating that the quadrupole moment more accurately represents the amplitude compared to higher-order moments. Notably, the 22 mode predominantly contributes to the radiation energy, rendering the contributions of other modes negligible. So here we only use the 22 mode of the quadrupole moment to calculate the radiation energy for the PN 22 mode. The PN initial parameters for 22 mode and orbital averaging are determined by fitting the corresponding NR waveform. Specifically, the PN orbital averaging parameters encompass $x_0$, ${e_t}_0$, while the initial parameters for the PN 22 mode involve $x_0$, ${e_t}_0$ and $l_0$. The former corresponds to the orbital average of the latter, or the average of $l$, ensuring that regardless of $l_0$ variations, the energy or angular momentum flux calculated by the former remains unchanged.
In FIG. \ref{FIG:4}, we maintain consistency in illustrating eccentricity by utilizing the initial eccentricity $e_0$ provided by RIT, and similarly for the temporal eccentricity ${e_t}_0$ of PN.
From FIG. \ref{FIG:4}, it is evident that the radiation energy $E_\text{rad}$ of the PN 22 mode and NR both exhibit oscillations, whereas the energy of the PN orbital averaging remains constant. The radiation energy of the PN 22 mode differs somewhat from that of the NR due to a larger amplitude error between the PN and NR compared to the frequency error, as indicated by the findings in Ref. \cite{Wang:2024jro}. This discrepancy causes the orbital average in FIG. \ref{FIG:4} to align closely with the NR, while the results of the PN 22 mode deviate. The absence of oscillation in the orbital average and the contrasting oscillations in NR and PN 22 mode depicted in FIG. \ref{FIG:4} highlight the impact of the mean anomaly $l_0$ on the radiated energy. This effect is smoothed out during the calculation of the orbital average energy flux. The presence of $l_0$ signifies an orbital influence that shapes the entire process of BBH merger in eccentric orbits.

\begin{figure*}[htbp!]
\centering
\includegraphics[width=16cm,height=5cm]{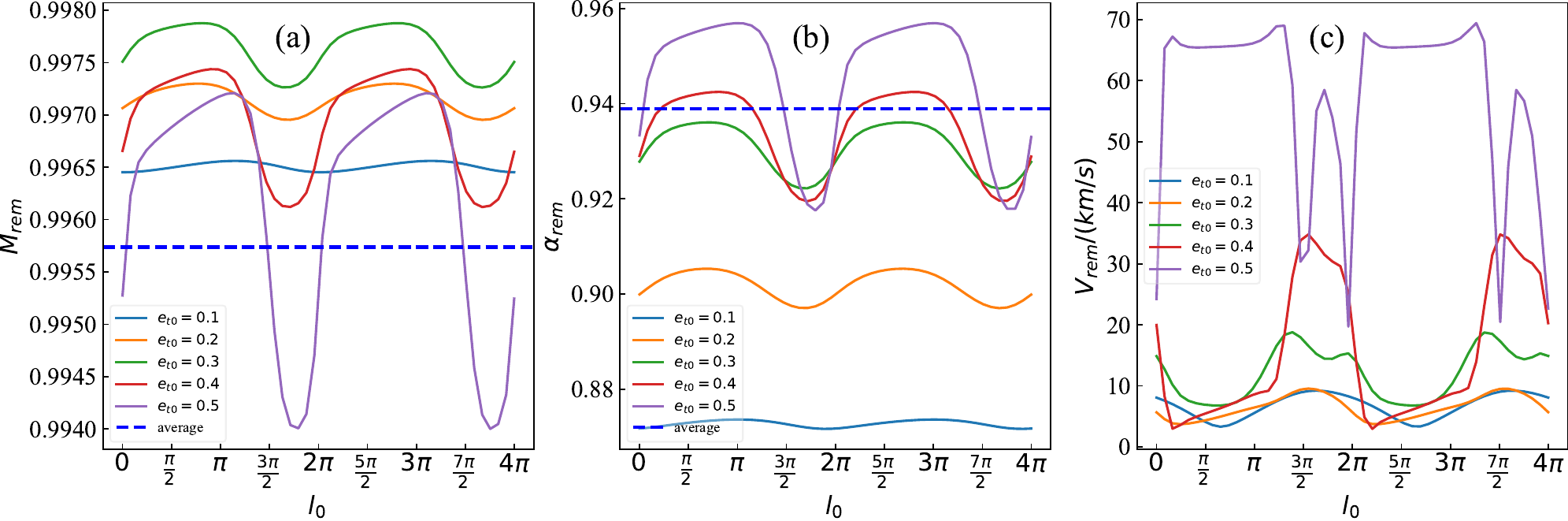}
\caption{\label{FIG:5}Relationship between mean anomaly $l_0$ and mass $M_\text{rem}$, spin $\alpha_\text{rem}$, and recoil velocity $V_\text{rem}$ of the remnant while keeping the initial eccentricity ${e_t}_0$ and parameter $x_0$ constant. For the same curve, ${e_t}_0$ is fixed, and for different curves, ${e_t}_0$ varies. The blue dashed line represents the orbital average for ${e_t}_0 = 0.5$.}
\end {figure*}

How does the variation of $l_0$ impact the dynamical quantities of mass, spin, and recoil velocity of the remnant? We investigate this issue through the perspective of PN theory for the inspiral phase. In FIG. \ref{FIG:5}, we illustrate the influence of changing $l_0$ on the mass $M_\text{rem}$, spin $\alpha_\text{rem}$, and recoil velocity $V_\text{rem}$ of the remnant while keeping the initial eccentricity ${e_t}_0$ and parameter $x_0$ constant. For symmetry, we set $q=0.5$ (as for $q=1$, the recoil velocity is zero), $x_0 =0.07$, and maintain a cutoff time of $-200M$. The initial eccentricity remains consistent for a single curve in FIG. \ref{FIG:5}. For $l_0$, a complete cycle spans $0$ to $2\pi$; for the sake of completeness, we extend this range to $0$ to $4\pi$. We strive to align the orbital averages of $M_\text{rem}$, $\alpha_\text{rem}$, and $V_\text{rem}$ as closely to the midpoint of the oscillations as possible, because any deviation would suggest a failure in orbital averaging, indicating proximity to the merger stage where PN methods falter. Here, we make an assumption that we exclude radiation during the merger and ringdown phases post $-200M$ as these are beyond PN control and do not impact problem elucidation.
We calculate $M_\text{rem}$ by $M_\text{rem}=M_{\mathrm{ADM}}-E_{\mathrm{rad}}$, with assuming $M_{\mathrm{ADM}}$ to be 1, that is, do not consider the effect of binding energy $E_{\mathrm{b}}$, which is very small and does not affect the final result. We calculate spin $\alpha_\text{rem}$ by $\alpha_\text{rem}=(L_0-L_z^{\mathrm{rad}})/M_{\text{rem}}^2$, with assuming $L_0$ to be 1, and due to symmetry, the radiation of angular momentum $L^{\mathrm{rad}} = L_z^{\mathrm{rad}}$ is concentrated in the $z$ direction (i.e. aligned with the direction of orbital angular momentum $L$), calculated by \cite{Ruiz:2007yx}:
\begin{equation}\label{eq:2}
L_{z}^{\mathrm{rad}} =\lim _{r \rightarrow \infty} \frac{ r^2}{16 \pi} \operatorname{Im}\left\{\sum_{\ell, m} m \int_{t_0}^t \left( h^{\ell m} \bar{\dot{h}}^{\ell m} \right)\mathrm{~d} t^{\prime}\right\},
\end{equation}
where $\bar{\dot{h}}^{\ell m}$ represents the complex conjugate of $\dot{h}^{\ell m}$.
The recoil velocity $V_\text{rem}$ is in the orbital plane can then be calculated by
$V_\text{rem}=|P^{\text{rad}}_{+}|/M_\text{rem}$, where

\begin{equation}\label{eq:3}
\begin{aligned}
P_{+}^{\mathrm{rad}} & =\lim _{r \rightarrow \infty} \frac{r^2}{8 \pi} \sum_{\ell, m} \int_{t_0}^t \mathrm{~d} t^{\prime} \dot{h}^{\ell m} 
\left(a_{\ell, m} \bar{\dot{h}}^{\ell, m+1} \right. \\
&\left. +b_{\ell,-m} \bar{\dot{h}}^{\ell-1, m+1} -b_{\ell+1, m+1} \bar{\dot{h}}_4^{\ell+1, m+1}\right),
\end{aligned}
\end{equation}
where the coefficients $a_{\ell, m}$, $b_{\ell,-m}$, and $b_{\ell+1, m+1}$ can be found in Ref. \cite{Ruiz:2007yx}. In FIG. \ref{FIG:5}, we only counted the 22 mode when calculating the mass $M_\text{rem}$, spin $\alpha_\text{rem}$, but included all the modes of $\ell \leq 4$ when calculating the recoil $V_\text{rem}$. The reason is that the harmonic mode has almost no effect on the calculation of mass and spin, but has a great impact on the recoil \cite{Radia:2021hjs}.

In FIG. \ref{FIG:5}, we present the average for ${e_t}_0 = 0.5$ for simplicity. The average of recoil velocity was not calculated due to the lack of corresponding calculated data.
From FIG. \ref{FIG:5}, it is evident that, for constant initial eccentricity, varying $l_0$ induces oscillations in these dynamic quantities; notably, higher initial eccentricities result in more pronounced oscillations. The figure clearly illustrates the amplitude of oscillation in dynamic quantities as eccentricity increases, which is consistent with the description in the FIG. \ref{FIG:3}. Specifically, for mass, spin, and recoil velocity, the amplitude of oscillations with ${e_t}_0 = 0.5$ are enhanced by approximately 10, 5, and 7 times, respectively, compared to those at ${e_t}_0 = 0.1$. It is foreseeable that for a circular orbit, i.e. eccentricity ${e_t}_0 = 0$, the oscillations will disappear.
It is important to note that the average values of these oscillations do not align perfectly, as we have not accounted for the merger and ringdown phases. Including these phases would yield average values that are more closely matched. Additionally, the recoil oscillations shown in panel (c) exhibit less regularity compared to those in panels (a) and (b). This irregularity stems from the recoil calculation formula (Eq. (\ref{eq:3})), which includes cross terms of harmonic modes, such as the interaction between the 22 and 33 modes. This interaction is the fundamental cause of the recoil anomalies discussed in Ref. \cite{Radia:2021hjs}. FIG. \ref{FIG:5} illustrates the impact of orbital effects during the inspiral phase; this effect would be significantly stronger if the merger and ringdown phases were also included.

FIG. \ref{FIG:3} reveals that the orbital effect manifests during the merger and ringdown phases, particularly in cases of high eccentricity. As the eccentricity increases, the oscillation becomes more pronounced. When only one complete orbit remains, this oscillation reaches its maximum amplitude. This is consistent with the principle illustrated in FIG. \ref{FIG:5}. This oscillation behavior captures the essence of how the orbital effect of eccentricity influences the merger dynamics, particularly in terms of peak amplitude.

However, the initial distance that NR can simulate is limited, which constrains the peak values of dynamic quantities observed at medium eccentricities (e.g., $e_0 = 0.3$ for $11.3M$ and $e_0 = 0.6$ for $24.6M$). In reality, higher initial eccentricities would lead to even greater enhancements, but it requires a larger initial distance for NR.

\noindent {\textbf{\textit{Discussion}}.}
The orbital effect in eccentric orbits is a unique and universal effect that influences the waveform and dynamics throughout the inspiral, merger, and ringdown phases. This effect is a key feature that differentiates eccentric orbits from circular ones. When constructing waveform templates or models for mass, spin, and recoil velocity of remnants in eccentric orbits, it is essential to account for this effect; neglecting it can introduce significant errors.
In future NR studies of BBH mergers in eccentric orbits, the orbital effect will be crucial for initial parameter setting, applicable to nonspinning, spin-aligned, and spinning-precessing BBH systems. In astrophysical environments, BBHs in eccentric orbits are prevalent. Under the influence of the orbital effect, some remnants from these mergers may experience changes in mass, angular momentum, and recoil velocity, which have important implications for understanding the dynamic formation channels of BBHs.

% \begin{acknowledgments}
\noindent {\textbf{\textit{Acknowledgments}}.}
The authors are very grateful to the RIT collaboration for the numerical simulation of eccentric BBH mergers, and thanks to Yan-Fang Huang, Xiaolin Liu, Zhou-Jian Cao for their helpful discussions. The computation is partially completed in the HPC Platform of Huazhong University of Science and Technology. The languages was polished by ChatGPT during the revision of the draft. This work is supported by the National Key R\&D Program of China (2021YFA0718504).
% \end{acknowledgments}

% \nocite{*}

\bibliography{ref}

%apsrev4-2.bst 2019-01-14 (MD) hand-edited version of apsrev4-1.bst
%Control: key (0)
%Control: author (8) initials jnrlst
%Control: editor formatted (1) identically to author
%Control: production of article title (0) allowed
%Control: page (0) single
%Control: year (1) truncated
%Control: production of eprint (0) enabled
\providecommand{\noopsort}[1]{}\providecommand{\singleletter}[1]{#1}%
\begin{thebibliography}{36}%
\makeatletter
\providecommand \@ifxundefined [1]{%
 \@ifx{#1\undefined}
}%
\providecommand \@ifnum [1]{%
 \ifnum #1\expandafter \@firstoftwo
 \else \expandafter \@secondoftwo
 \fi
}%
\providecommand \@ifx [1]{%
 \ifx #1\expandafter \@firstoftwo
 \else \expandafter \@secondoftwo
 \fi
}%
\providecommand \natexlab [1]{#1}%
\providecommand \enquote  [1]{``#1''}%
\providecommand \bibnamefont  [1]{#1}%
\providecommand \bibfnamefont [1]{#1}%
\providecommand \citenamefont [1]{#1}%
\providecommand \href@noop [0]{\@secondoftwo}%
\providecommand \href [0]{\begingroup \@sanitize@url \@href}%
\providecommand \@href[1]{\@@startlink{#1}\@@href}%
\providecommand \@@href[1]{\endgroup#1\@@endlink}%
\providecommand \@sanitize@url [0]{\catcode `\\12\catcode `\$12\catcode `\&12\catcode `\#12\catcode `\^12\catcode `\_12\catcode `\%12\relax}%
\providecommand \@@startlink[1]{}%
\providecommand \@@endlink[0]{}%
\providecommand \url  [0]{\begingroup\@sanitize@url \@url }%
\providecommand \@url [1]{\endgroup\@href {#1}{\urlprefix }}%
\providecommand \urlprefix  [0]{URL }%
\providecommand \Eprint [0]{\href }%
\providecommand \doibase [0]{https://doi.org/}%
\providecommand \selectlanguage [0]{\@gobble}%
\providecommand \bibinfo  [0]{\@secondoftwo}%
\providecommand \bibfield  [0]{\@secondoftwo}%
\providecommand \translation [1]{[#1]}%
\providecommand \BibitemOpen [0]{}%
\providecommand \bibitemStop [0]{}%
\providecommand \bibitemNoStop [0]{.\EOS\space}%
\providecommand \EOS [0]{\spacefactor3000\relax}%
\providecommand \BibitemShut  [1]{\csname bibitem#1\endcsname}%
\let\auto@bib@innerbib\@empty
%</preamble>
\bibitem [{\citenamefont {Aasi}\ \emph {et~al.}(2015)\citenamefont {Aasi} \emph {et~al.}}]{LIGOScientific:2014qfs}%
  \BibitemOpen
  \bibfield  {author} {\bibinfo {author} {\bibfnamefont {J.}~\bibnamefont {Aasi}} \emph {et~al.} (\bibinfo {collaboration} {LIGO Scientific, VIRGO}),\ }\bibfield  {title} {\bibinfo {title} {{Characterization of the LIGO detectors during their sixth science run}},\ }\href {https://doi.org/10.1088/0264-9381/32/11/115012} {\bibfield  {journal} {\bibinfo  {journal} {Class. Quant. Grav.}\ }\textbf {\bibinfo {volume} {32}},\ \bibinfo {pages} {115012} (\bibinfo {year} {2015})},\ \Eprint {https://arxiv.org/abs/1410.7764} {arXiv:1410.7764 [gr-qc]} \BibitemShut {NoStop}%
\bibitem [{\citenamefont {Acernese}\ \emph {et~al.}(2015)\citenamefont {Acernese} \emph {et~al.}}]{VIRGO:2014yos}%
  \BibitemOpen
  \bibfield  {author} {\bibinfo {author} {\bibfnamefont {F.}~\bibnamefont {Acernese}} \emph {et~al.} (\bibinfo {collaboration} {VIRGO}),\ }\bibfield  {title} {\bibinfo {title} {{Advanced Virgo: a second-generation interferometric gravitational wave detector}},\ }\href {https://doi.org/10.1088/0264-9381/32/2/024001} {\bibfield  {journal} {\bibinfo  {journal} {Class. Quant. Grav.}\ }\textbf {\bibinfo {volume} {32}},\ \bibinfo {pages} {024001} (\bibinfo {year} {2015})},\ \Eprint {https://arxiv.org/abs/1408.3978} {arXiv:1408.3978 [gr-qc]} \BibitemShut {NoStop}%
\bibitem [{\citenamefont {Akutsu}\ \emph {et~al.}(2019)\citenamefont {Akutsu} \emph {et~al.}}]{KAGRA:2018plz}%
  \BibitemOpen
  \bibfield  {author} {\bibinfo {author} {\bibfnamefont {T.}~\bibnamefont {Akutsu}} \emph {et~al.} (\bibinfo {collaboration} {KAGRA}),\ }\bibfield  {title} {\bibinfo {title} {{KAGRA: 2.5 Generation Interferometric Gravitational Wave Detector}},\ }\href {https://doi.org/10.1038/s41550-018-0658-y} {\bibfield  {journal} {\bibinfo  {journal} {Nature Astron.}\ }\textbf {\bibinfo {volume} {3}},\ \bibinfo {pages} {35} (\bibinfo {year} {2019})},\ \Eprint {https://arxiv.org/abs/1811.08079} {arXiv:1811.08079 [gr-qc]} \BibitemShut {NoStop}%
\bibitem [{\citenamefont {Peters}\ and\ \citenamefont {Mathews}(1963)}]{Peters:1963ux}%
  \BibitemOpen
  \bibfield  {author} {\bibinfo {author} {\bibfnamefont {P.~C.}\ \bibnamefont {Peters}}\ and\ \bibinfo {author} {\bibfnamefont {J.}~\bibnamefont {Mathews}},\ }\bibfield  {title} {\bibinfo {title} {{Gravitational radiation from point masses in a Keplerian orbit}},\ }\href {https://doi.org/10.1103/PhysRev.131.435} {\bibfield  {journal} {\bibinfo  {journal} {Phys. Rev.}\ }\textbf {\bibinfo {volume} {131}},\ \bibinfo {pages} {435} (\bibinfo {year} {1963})}\BibitemShut {NoStop}%
\bibitem [{\citenamefont {Peters}(1964)}]{Peters:1964zz}%
  \BibitemOpen
  \bibfield  {author} {\bibinfo {author} {\bibfnamefont {P.~C.}\ \bibnamefont {Peters}},\ }\bibfield  {title} {\bibinfo {title} {{Gravitational Radiation and the Motion of Two Point Masses}},\ }\href {https://doi.org/10.1103/PhysRev.136.B1224} {\bibfield  {journal} {\bibinfo  {journal} {Phys. Rev.}\ }\textbf {\bibinfo {volume} {136}},\ \bibinfo {pages} {B1224} (\bibinfo {year} {1964})}\BibitemShut {NoStop}%
\bibitem [{\citenamefont {Miller}\ and\ \citenamefont {Hamilton}(2002)}]{Miller:2002pg}%
  \BibitemOpen
  \bibfield  {author} {\bibinfo {author} {\bibfnamefont {M.~C.}\ \bibnamefont {Miller}}\ and\ \bibinfo {author} {\bibfnamefont {D.~P.}\ \bibnamefont {Hamilton}},\ }\bibfield  {title} {\bibinfo {title} {{Four-body effects in globular cluster black hole coalescence}},\ }\href {https://doi.org/10.1086/341788} {\bibfield  {journal} {\bibinfo  {journal} {Astrophys. J.}\ }\textbf {\bibinfo {volume} {576}},\ \bibinfo {pages} {894} (\bibinfo {year} {2002})},\ \Eprint {https://arxiv.org/abs/astro-ph/0202298} {arXiv:astro-ph/0202298} \BibitemShut {NoStop}%
\bibitem [{\citenamefont {Rodriguez}\ \emph {et~al.}(2016)\citenamefont {Rodriguez}, \citenamefont {Chatterjee},\ and\ \citenamefont {Rasio}}]{Rodriguez:2016kxx}%
  \BibitemOpen
  \bibfield  {author} {\bibinfo {author} {\bibfnamefont {C.~L.}\ \bibnamefont {Rodriguez}}, \bibinfo {author} {\bibfnamefont {S.}~\bibnamefont {Chatterjee}},\ and\ \bibinfo {author} {\bibfnamefont {F.~A.}\ \bibnamefont {Rasio}},\ }\bibfield  {title} {\bibinfo {title} {{Binary Black Hole Mergers from Globular Clusters: Masses, Merger Rates, and the Impact of Stellar Evolution}},\ }\href {https://doi.org/10.1103/PhysRevD.93.084029} {\bibfield  {journal} {\bibinfo  {journal} {Phys. Rev. D}\ }\textbf {\bibinfo {volume} {93}},\ \bibinfo {pages} {084029} (\bibinfo {year} {2016})},\ \Eprint {https://arxiv.org/abs/1602.02444} {arXiv:1602.02444 [astro-ph.HE]} \BibitemShut {NoStop}%
\bibitem [{\citenamefont {Gond\'an}\ and\ \citenamefont {Kocsis}(2021)}]{Gondan:2020svr}%
  \BibitemOpen
  \bibfield  {author} {\bibinfo {author} {\bibfnamefont {L.}~\bibnamefont {Gond\'an}}\ and\ \bibinfo {author} {\bibfnamefont {B.}~\bibnamefont {Kocsis}},\ }\bibfield  {title} {\bibinfo {title} {{High eccentricities and high masses characterize gravitational-wave captures in galactic nuclei as seen by Earth-based detectors}},\ }\href {https://doi.org/10.1093/mnras/stab1722} {\bibfield  {journal} {\bibinfo  {journal} {Mon. Not. Roy. Astron. Soc.}\ }\textbf {\bibinfo {volume} {506}},\ \bibinfo {pages} {1665} (\bibinfo {year} {2021})},\ \Eprint {https://arxiv.org/abs/2011.02507} {arXiv:2011.02507 [astro-ph.HE]} \BibitemShut {NoStop}%
\bibitem [{\citenamefont {Gultekin}\ \emph {et~al.}(2006)\citenamefont {Gultekin}, \citenamefont {Coleman~Miller},\ and\ \citenamefont {Hamilton}}]{Gultekin:2005fd}%
  \BibitemOpen
  \bibfield  {author} {\bibinfo {author} {\bibfnamefont {K.}~\bibnamefont {Gultekin}}, \bibinfo {author} {\bibfnamefont {M.}~\bibnamefont {Coleman~Miller}},\ and\ \bibinfo {author} {\bibfnamefont {D.~P.}\ \bibnamefont {Hamilton}},\ }\bibfield  {title} {\bibinfo {title} {{Three-body dynamics with gravitational wave emission}},\ }\href {https://doi.org/10.1086/499917} {\bibfield  {journal} {\bibinfo  {journal} {Astrophys. J.}\ }\textbf {\bibinfo {volume} {640}},\ \bibinfo {pages} {156} (\bibinfo {year} {2006})},\ \Eprint {https://arxiv.org/abs/astro-ph/0509885} {arXiv:astro-ph/0509885} \BibitemShut {NoStop}%
\bibitem [{\citenamefont {Samsing}\ \emph {et~al.}(2014)\citenamefont {Samsing}, \citenamefont {MacLeod},\ and\ \citenamefont {Ramirez-Ruiz}}]{Samsing:2013kua}%
  \BibitemOpen
  \bibfield  {author} {\bibinfo {author} {\bibfnamefont {J.}~\bibnamefont {Samsing}}, \bibinfo {author} {\bibfnamefont {M.}~\bibnamefont {MacLeod}},\ and\ \bibinfo {author} {\bibfnamefont {E.}~\bibnamefont {Ramirez-Ruiz}},\ }\bibfield  {title} {\bibinfo {title} {{The Formation of Eccentric Compact Binary Inspirals and the Role of Gravitational Wave Emission in Binary-Single Stellar Encounters}},\ }\href {https://doi.org/10.1088/0004-637X/784/1/71} {\bibfield  {journal} {\bibinfo  {journal} {Astrophys. J.}\ }\textbf {\bibinfo {volume} {784}},\ \bibinfo {pages} {71} (\bibinfo {year} {2014})},\ \Eprint {https://arxiv.org/abs/1308.2964} {arXiv:1308.2964 [astro-ph.HE]} \BibitemShut {NoStop}%
\bibitem [{\citenamefont {Samsing}\ \emph {et~al.}(2018)\citenamefont {Samsing}, \citenamefont {Askar},\ and\ \citenamefont {Giersz}}]{Samsing:2017oij}%
  \BibitemOpen
  \bibfield  {author} {\bibinfo {author} {\bibfnamefont {J.}~\bibnamefont {Samsing}}, \bibinfo {author} {\bibfnamefont {A.}~\bibnamefont {Askar}},\ and\ \bibinfo {author} {\bibfnamefont {M.}~\bibnamefont {Giersz}},\ }\bibfield  {title} {\bibinfo {title} {{MOCCA-SURVEY Database. I. Eccentric Black Hole Mergers during Binary\textendash{}Single Interactions in Globular Clusters}},\ }\href {https://doi.org/10.3847/1538-4357/aaab52} {\bibfield  {journal} {\bibinfo  {journal} {Astrophys. J.}\ }\textbf {\bibinfo {volume} {855}},\ \bibinfo {pages} {124} (\bibinfo {year} {2018})},\ \Eprint {https://arxiv.org/abs/1712.06186} {arXiv:1712.06186 [astro-ph.HE]} \BibitemShut {NoStop}%
\bibitem [{\citenamefont {Abbott}\ \emph {et~al.}(2020)\citenamefont {Abbott} \emph {et~al.}}]{LIGOScientific:2020iuh}%
  \BibitemOpen
  \bibfield  {author} {\bibinfo {author} {\bibfnamefont {R.}~\bibnamefont {Abbott}} \emph {et~al.} (\bibinfo {collaboration} {LIGO Scientific, Virgo}),\ }\bibfield  {title} {\bibinfo {title} {{GW190521: A Binary Black Hole Merger with a Total Mass of $150 M_{\odot}$}},\ }\href {https://doi.org/10.1103/PhysRevLett.125.101102} {\bibfield  {journal} {\bibinfo  {journal} {Phys. Rev. Lett.}\ }\textbf {\bibinfo {volume} {125}},\ \bibinfo {pages} {101102} (\bibinfo {year} {2020})},\ \Eprint {https://arxiv.org/abs/2009.01075} {arXiv:2009.01075 [gr-qc]} \BibitemShut {NoStop}%
\bibitem [{\citenamefont {Gayathri}\ \emph {et~al.}(2022)\citenamefont {Gayathri}, \citenamefont {Healy}, \citenamefont {Lange}, \citenamefont {O'Brien}, \citenamefont {Szczepanczyk}, \citenamefont {Bartos}, \citenamefont {Campanelli}, \citenamefont {Klimenko}, \citenamefont {Lousto},\ and\ \citenamefont {O'Shaughnessy}}]{Gayathri:2020coq}%
  \BibitemOpen
  \bibfield  {author} {\bibinfo {author} {\bibfnamefont {V.}~\bibnamefont {Gayathri}}, \bibinfo {author} {\bibfnamefont {J.}~\bibnamefont {Healy}}, \bibinfo {author} {\bibfnamefont {J.}~\bibnamefont {Lange}}, \bibinfo {author} {\bibfnamefont {B.}~\bibnamefont {O'Brien}}, \bibinfo {author} {\bibfnamefont {M.}~\bibnamefont {Szczepanczyk}}, \bibinfo {author} {\bibfnamefont {I.}~\bibnamefont {Bartos}}, \bibinfo {author} {\bibfnamefont {M.}~\bibnamefont {Campanelli}}, \bibinfo {author} {\bibfnamefont {S.}~\bibnamefont {Klimenko}}, \bibinfo {author} {\bibfnamefont {C.~O.}\ \bibnamefont {Lousto}},\ and\ \bibinfo {author} {\bibfnamefont {R.}~\bibnamefont {O'Shaughnessy}},\ }\bibfield  {title} {\bibinfo {title} {{Eccentricity estimate for black hole mergers with numerical relativity simulations}},\ }\href {https://doi.org/10.1038/s41550-021-01568-w} {\bibfield  {journal} {\bibinfo  {journal} {Nature Astron.}\ }\textbf {\bibinfo {volume} {6}},\ \bibinfo {pages} {344} (\bibinfo {year} {2022})},\ \Eprint
  {https://arxiv.org/abs/2009.05461} {arXiv:2009.05461 [astro-ph.HE]} \BibitemShut {NoStop}%
\bibitem [{\citenamefont {Blanchet}(2014)}]{Blanchet:2013haa}%
  \BibitemOpen
  \bibfield  {author} {\bibinfo {author} {\bibfnamefont {L.}~\bibnamefont {Blanchet}},\ }\bibfield  {title} {\bibinfo {title} {{Post-Newtonian Theory for Gravitational Waves}},\ }\href {https://doi.org/10.12942/lrr-2014-2} {\bibfield  {journal} {\bibinfo  {journal} {Living Rev. Rel.}\ }\textbf {\bibinfo {volume} {17}},\ \bibinfo {pages} {2} (\bibinfo {year} {2014})},\ \Eprint {https://arxiv.org/abs/1310.1528} {arXiv:1310.1528 [gr-qc]} \BibitemShut {NoStop}%
\bibitem [{\citenamefont {Sperhake}\ \emph {et~al.}(2008)\citenamefont {Sperhake}, \citenamefont {Berti}, \citenamefont {Cardoso}, \citenamefont {Gonzalez}, \citenamefont {Bruegmann},\ and\ \citenamefont {Ansorg}}]{Sperhake:2007gu}%
  \BibitemOpen
  \bibfield  {author} {\bibinfo {author} {\bibfnamefont {U.}~\bibnamefont {Sperhake}}, \bibinfo {author} {\bibfnamefont {E.}~\bibnamefont {Berti}}, \bibinfo {author} {\bibfnamefont {V.}~\bibnamefont {Cardoso}}, \bibinfo {author} {\bibfnamefont {J.~A.}\ \bibnamefont {Gonzalez}}, \bibinfo {author} {\bibfnamefont {B.}~\bibnamefont {Bruegmann}},\ and\ \bibinfo {author} {\bibfnamefont {M.}~\bibnamefont {Ansorg}},\ }\bibfield  {title} {\bibinfo {title} {{Eccentric binary black-hole mergers: The Transition from inspiral to plunge in general relativity}},\ }\href {https://doi.org/10.1103/PhysRevD.78.064069} {\bibfield  {journal} {\bibinfo  {journal} {Phys. Rev. D}\ }\textbf {\bibinfo {volume} {78}},\ \bibinfo {pages} {064069} (\bibinfo {year} {2008})},\ \Eprint {https://arxiv.org/abs/0710.3823} {arXiv:0710.3823 [gr-qc]} \BibitemShut {NoStop}%
\bibitem [{\citenamefont {Hinder}\ \emph {et~al.}(2008)\citenamefont {Hinder}, \citenamefont {Vaishnav}, \citenamefont {Herrmann}, \citenamefont {Shoemaker},\ and\ \citenamefont {Laguna}}]{Hinder:2007qu}%
  \BibitemOpen
  \bibfield  {author} {\bibinfo {author} {\bibfnamefont {I.}~\bibnamefont {Hinder}}, \bibinfo {author} {\bibfnamefont {B.}~\bibnamefont {Vaishnav}}, \bibinfo {author} {\bibfnamefont {F.}~\bibnamefont {Herrmann}}, \bibinfo {author} {\bibfnamefont {D.}~\bibnamefont {Shoemaker}},\ and\ \bibinfo {author} {\bibfnamefont {P.}~\bibnamefont {Laguna}},\ }\bibfield  {title} {\bibinfo {title} {{Universality and final spin in eccentric binary black hole inspirals}},\ }\href {https://doi.org/10.1103/PhysRevD.77.081502} {\bibfield  {journal} {\bibinfo  {journal} {Phys. Rev. D}\ }\textbf {\bibinfo {volume} {77}},\ \bibinfo {pages} {081502} (\bibinfo {year} {2008})},\ \Eprint {https://arxiv.org/abs/0710.5167} {arXiv:0710.5167 [gr-qc]} \BibitemShut {NoStop}%
\bibitem [{\citenamefont {Huerta}\ \emph {et~al.}(2019)\citenamefont {Huerta} \emph {et~al.}}]{Huerta:2019oxn}%
  \BibitemOpen
  \bibfield  {author} {\bibinfo {author} {\bibfnamefont {E.~A.}\ \bibnamefont {Huerta}} \emph {et~al.},\ }\bibfield  {title} {\bibinfo {title} {{Physics of eccentric binary black hole mergers: A numerical relativity perspective}},\ }\href {https://doi.org/10.1103/PhysRevD.100.064003} {\bibfield  {journal} {\bibinfo  {journal} {Phys. Rev. D}\ }\textbf {\bibinfo {volume} {100}},\ \bibinfo {pages} {064003} (\bibinfo {year} {2019})},\ \Eprint {https://arxiv.org/abs/1901.07038} {arXiv:1901.07038 [gr-qc]} \BibitemShut {NoStop}%
\bibitem [{\citenamefont {Sperhake}\ \emph {et~al.}(2020)\citenamefont {Sperhake}, \citenamefont {Rosca-Mead}, \citenamefont {Gerosa},\ and\ \citenamefont {Berti}}]{Sperhake:2019wwo}%
  \BibitemOpen
  \bibfield  {author} {\bibinfo {author} {\bibfnamefont {U.}~\bibnamefont {Sperhake}}, \bibinfo {author} {\bibfnamefont {R.}~\bibnamefont {Rosca-Mead}}, \bibinfo {author} {\bibfnamefont {D.}~\bibnamefont {Gerosa}},\ and\ \bibinfo {author} {\bibfnamefont {E.}~\bibnamefont {Berti}},\ }\bibfield  {title} {\bibinfo {title} {{Amplification of superkicks in black-hole binaries through orbital eccentricity}},\ }\href {https://doi.org/10.1103/PhysRevD.101.024044} {\bibfield  {journal} {\bibinfo  {journal} {Phys. Rev. D}\ }\textbf {\bibinfo {volume} {101}},\ \bibinfo {pages} {024044} (\bibinfo {year} {2020})},\ \Eprint {https://arxiv.org/abs/1910.01598} {arXiv:1910.01598 [gr-qc]} \BibitemShut {NoStop}%
\bibitem [{\citenamefont {Radia}\ \emph {et~al.}(2021)\citenamefont {Radia}, \citenamefont {Sperhake}, \citenamefont {Berti},\ and\ \citenamefont {Croft}}]{Radia:2021hjs}%
  \BibitemOpen
  \bibfield  {author} {\bibinfo {author} {\bibfnamefont {M.}~\bibnamefont {Radia}}, \bibinfo {author} {\bibfnamefont {U.}~\bibnamefont {Sperhake}}, \bibinfo {author} {\bibfnamefont {E.}~\bibnamefont {Berti}},\ and\ \bibinfo {author} {\bibfnamefont {R.}~\bibnamefont {Croft}},\ }\bibfield  {title} {\bibinfo {title} {{Anomalies in the gravitational recoil of eccentric black-hole mergers with unequal mass ratios}},\ }\href {https://doi.org/10.1103/PhysRevD.103.104006} {\bibfield  {journal} {\bibinfo  {journal} {Phys. Rev. D}\ }\textbf {\bibinfo {volume} {103}},\ \bibinfo {pages} {104006} (\bibinfo {year} {2021})},\ \Eprint {https://arxiv.org/abs/2101.11015} {arXiv:2101.11015 [gr-qc]} \BibitemShut {NoStop}%
\bibitem [{\citenamefont {Healy}\ and\ \citenamefont {Lousto}(2022)}]{Healy:2022wdn}%
  \BibitemOpen
  \bibfield  {author} {\bibinfo {author} {\bibfnamefont {J.}~\bibnamefont {Healy}}\ and\ \bibinfo {author} {\bibfnamefont {C.~O.}\ \bibnamefont {Lousto}},\ }\bibfield  {title} {\bibinfo {title} {{Fourth RIT binary black hole simulations catalog: Extension to eccentric orbits}},\ }\href {https://doi.org/10.1103/PhysRevD.105.124010} {\bibfield  {journal} {\bibinfo  {journal} {Phys. Rev. D}\ }\textbf {\bibinfo {volume} {105}},\ \bibinfo {pages} {124010} (\bibinfo {year} {2022})},\ \Eprint {https://arxiv.org/abs/2202.00018} {arXiv:2202.00018 [gr-qc]} \BibitemShut {NoStop}%
\bibitem [{\citenamefont {Wang}\ \emph {et~al.}(2024{\natexlab{a}})\citenamefont {Wang}, \citenamefont {Zou}, \citenamefont {Wu}, \citenamefont {Liu},\ and\ \citenamefont {Liu}}]{Wang:2023vka}%
  \BibitemOpen
  \bibfield  {author} {\bibinfo {author} {\bibfnamefont {H.}~\bibnamefont {Wang}}, \bibinfo {author} {\bibfnamefont {Y.-C.}\ \bibnamefont {Zou}}, \bibinfo {author} {\bibfnamefont {Q.-W.}\ \bibnamefont {Wu}}, \bibinfo {author} {\bibfnamefont {Y.}~\bibnamefont {Liu}},\ and\ \bibinfo {author} {\bibfnamefont {X.}~\bibnamefont {Liu}},\ }\bibfield  {title} {\bibinfo {title} {{Characterizing the effect of eccentricity on the dynamics of binary black hole mergers in numerical relativity}},\ }\href {https://doi.org/10.1103/PhysRevD.109.084063} {\bibfield  {journal} {\bibinfo  {journal} {Phys. Rev. D}\ }\textbf {\bibinfo {volume} {109}},\ \bibinfo {pages} {084063} (\bibinfo {year} {2024}{\natexlab{a}})},\ \Eprint {https://arxiv.org/abs/2310.04777} {arXiv:2310.04777 [gr-qc]} \BibitemShut {NoStop}%
\bibitem [{\citenamefont {Ruiz}\ \emph {et~al.}(2008)\citenamefont {Ruiz}, \citenamefont {Takahashi}, \citenamefont {Alcubierre},\ and\ \citenamefont {Nunez}}]{Ruiz:2007yx}%
  \BibitemOpen
  \bibfield  {author} {\bibinfo {author} {\bibfnamefont {M.}~\bibnamefont {Ruiz}}, \bibinfo {author} {\bibfnamefont {R.}~\bibnamefont {Takahashi}}, \bibinfo {author} {\bibfnamefont {M.}~\bibnamefont {Alcubierre}},\ and\ \bibinfo {author} {\bibfnamefont {D.}~\bibnamefont {Nunez}},\ }\bibfield  {title} {\bibinfo {title} {{Multipole expansions for energy and momenta carried by gravitational waves}},\ }\href {https://doi.org/10.1007/s10714-007-0570-8} {\bibfield  {journal} {\bibinfo  {journal} {Gen. Rel. Grav.}\ }\textbf {\bibinfo {volume} {40}},\ \bibinfo {pages} {2467} (\bibinfo {year} {2008})},\ \Eprint {https://arxiv.org/abs/0707.4654} {arXiv:0707.4654 [gr-qc]} \BibitemShut {NoStop}%
\bibitem [{\citenamefont {Wang}\ \emph {et~al.}(2024{\natexlab{b}})\citenamefont {Wang}, \citenamefont {Zou}, \citenamefont {Wu}, \citenamefont {Liu},\ and\ \citenamefont {Li}}]{Wang:2024jro}%
  \BibitemOpen
  \bibfield  {author} {\bibinfo {author} {\bibfnamefont {H.}~\bibnamefont {Wang}}, \bibinfo {author} {\bibfnamefont {Y.-C.}\ \bibnamefont {Zou}}, \bibinfo {author} {\bibfnamefont {Q.-W.}\ \bibnamefont {Wu}}, \bibinfo {author} {\bibfnamefont {X.}~\bibnamefont {Liu}},\ and\ \bibinfo {author} {\bibfnamefont {Z.}~\bibnamefont {Li}},\ }\bibfield  {title} {\bibinfo {title} {{A complete waveform comparison of post-Newtonian and numerical relativity in eccentric orbits}},\ }\href@noop {} {\  (\bibinfo {year} {2024}{\natexlab{b}})},\ \Eprint {https://arxiv.org/abs/2409.17636} {arXiv:2409.17636 [gr-qc]} \BibitemShut {NoStop}%
\bibitem [{\citenamefont {Healy}\ \emph {et~al.}(2017)\citenamefont {Healy}, \citenamefont {Lousto}, \citenamefont {Zlochower},\ and\ \citenamefont {Campanelli}}]{Healy:2017psd}%
  \BibitemOpen
  \bibfield  {author} {\bibinfo {author} {\bibfnamefont {J.}~\bibnamefont {Healy}}, \bibinfo {author} {\bibfnamefont {C.~O.}\ \bibnamefont {Lousto}}, \bibinfo {author} {\bibfnamefont {Y.}~\bibnamefont {Zlochower}},\ and\ \bibinfo {author} {\bibfnamefont {M.}~\bibnamefont {Campanelli}},\ }\bibfield  {title} {\bibinfo {title} {{The RIT binary black hole simulations catalog}},\ }\href {https://doi.org/10.1088/1361-6382/aa91b1} {\bibfield  {journal} {\bibinfo  {journal} {Class. Quant. Grav.}\ }\textbf {\bibinfo {volume} {34}},\ \bibinfo {pages} {224001} (\bibinfo {year} {2017})},\ \Eprint {https://arxiv.org/abs/1703.03423} {arXiv:1703.03423 [gr-qc]} \BibitemShut {NoStop}%
\bibitem [{\citenamefont {Healy}\ \emph {et~al.}(2019)\citenamefont {Healy}, \citenamefont {Lousto}, \citenamefont {Lange}, \citenamefont {O'Shaughnessy}, \citenamefont {Zlochower},\ and\ \citenamefont {Campanelli}}]{Healy:2019jyf}%
  \BibitemOpen
  \bibfield  {author} {\bibinfo {author} {\bibfnamefont {J.}~\bibnamefont {Healy}}, \bibinfo {author} {\bibfnamefont {C.~O.}\ \bibnamefont {Lousto}}, \bibinfo {author} {\bibfnamefont {J.}~\bibnamefont {Lange}}, \bibinfo {author} {\bibfnamefont {R.}~\bibnamefont {O'Shaughnessy}}, \bibinfo {author} {\bibfnamefont {Y.}~\bibnamefont {Zlochower}},\ and\ \bibinfo {author} {\bibfnamefont {M.}~\bibnamefont {Campanelli}},\ }\bibfield  {title} {\bibinfo {title} {{Second RIT binary black hole simulations catalog and its application to gravitational waves parameter estimation}},\ }\href {https://doi.org/10.1103/PhysRevD.100.024021} {\bibfield  {journal} {\bibinfo  {journal} {Phys. Rev. D}\ }\textbf {\bibinfo {volume} {100}},\ \bibinfo {pages} {024021} (\bibinfo {year} {2019})},\ \Eprint {https://arxiv.org/abs/1901.02553} {arXiv:1901.02553 [gr-qc]} \BibitemShut {NoStop}%
\bibitem [{\citenamefont {Healy}\ and\ \citenamefont {Lousto}(2020)}]{Healy:2020vre}%
  \BibitemOpen
  \bibfield  {author} {\bibinfo {author} {\bibfnamefont {J.}~\bibnamefont {Healy}}\ and\ \bibinfo {author} {\bibfnamefont {C.~O.}\ \bibnamefont {Lousto}},\ }\bibfield  {title} {\bibinfo {title} {{Third RIT binary black hole simulations catalog}},\ }\href {https://doi.org/10.1103/PhysRevD.102.104018} {\bibfield  {journal} {\bibinfo  {journal} {Phys. Rev. D}\ }\textbf {\bibinfo {volume} {102}},\ \bibinfo {pages} {104018} (\bibinfo {year} {2020})},\ \Eprint {https://arxiv.org/abs/2007.07910} {arXiv:2007.07910 [gr-qc]} \BibitemShut {NoStop}%
\bibitem [{\citenamefont {Mroue}\ \emph {et~al.}(2013)\citenamefont {Mroue} \emph {et~al.}}]{Mroue:2013xna}%
  \BibitemOpen
  \bibfield  {author} {\bibinfo {author} {\bibfnamefont {A.~H.}\ \bibnamefont {Mroue}} \emph {et~al.},\ }\bibfield  {title} {\bibinfo {title} {{Catalog of 174 Binary Black Hole Simulations for Gravitational Wave Astronomy}},\ }\href {https://doi.org/10.1103/PhysRevLett.111.241104} {\bibfield  {journal} {\bibinfo  {journal} {Phys. Rev. Lett.}\ }\textbf {\bibinfo {volume} {111}},\ \bibinfo {pages} {241104} (\bibinfo {year} {2013})},\ \Eprint {https://arxiv.org/abs/1304.6077} {arXiv:1304.6077 [gr-qc]} \BibitemShut {NoStop}%
\bibitem [{\citenamefont {Boyle}\ \emph {et~al.}(2019)\citenamefont {Boyle} \emph {et~al.}}]{Boyle:2019kee}%
  \BibitemOpen
  \bibfield  {author} {\bibinfo {author} {\bibfnamefont {M.}~\bibnamefont {Boyle}} \emph {et~al.},\ }\bibfield  {title} {\bibinfo {title} {{The SXS Collaboration catalog of binary black hole simulations}},\ }\href {https://doi.org/10.1088/1361-6382/ab34e2} {\bibfield  {journal} {\bibinfo  {journal} {Class. Quant. Grav.}\ }\textbf {\bibinfo {volume} {36}},\ \bibinfo {pages} {195006} (\bibinfo {year} {2019})},\ \Eprint {https://arxiv.org/abs/1904.04831} {arXiv:1904.04831 [gr-qc]} \BibitemShut {NoStop}%
\bibitem [{\citenamefont {Hinder}\ \emph {et~al.}(2010)\citenamefont {Hinder}, \citenamefont {Herrmann}, \citenamefont {Laguna},\ and\ \citenamefont {Shoemaker}}]{Hinder:2008kv}%
  \BibitemOpen
  \bibfield  {author} {\bibinfo {author} {\bibfnamefont {I.}~\bibnamefont {Hinder}}, \bibinfo {author} {\bibfnamefont {F.}~\bibnamefont {Herrmann}}, \bibinfo {author} {\bibfnamefont {P.}~\bibnamefont {Laguna}},\ and\ \bibinfo {author} {\bibfnamefont {D.}~\bibnamefont {Shoemaker}},\ }\bibfield  {title} {\bibinfo {title} {{Comparisons of eccentric binary black hole simulations with post-Newtonian models}},\ }\href {https://doi.org/10.1103/PhysRevD.82.024033} {\bibfield  {journal} {\bibinfo  {journal} {Phys. Rev. D}\ }\textbf {\bibinfo {volume} {82}},\ \bibinfo {pages} {024033} (\bibinfo {year} {2010})},\ \Eprint {https://arxiv.org/abs/0806.1037} {arXiv:0806.1037 [gr-qc]} \BibitemShut {NoStop}%
\bibitem [{\citenamefont {Memmesheimer}\ \emph {et~al.}(2004)\citenamefont {Memmesheimer}, \citenamefont {Gopakumar},\ and\ \citenamefont {Schaefer}}]{Memmesheimer:2004cv}%
  \BibitemOpen
  \bibfield  {author} {\bibinfo {author} {\bibfnamefont {R.-M.}\ \bibnamefont {Memmesheimer}}, \bibinfo {author} {\bibfnamefont {A.}~\bibnamefont {Gopakumar}},\ and\ \bibinfo {author} {\bibfnamefont {G.}~\bibnamefont {Schaefer}},\ }\bibfield  {title} {\bibinfo {title} {{Third post-Newtonian accurate generalized quasi-Keplerian parametrization for compact binaries in eccentric orbits}},\ }\href {https://doi.org/10.1103/PhysRevD.70.104011} {\bibfield  {journal} {\bibinfo  {journal} {Phys. Rev. D}\ }\textbf {\bibinfo {volume} {70}},\ \bibinfo {pages} {104011} (\bibinfo {year} {2004})},\ \Eprint {https://arxiv.org/abs/gr-qc/0407049} {arXiv:gr-qc/0407049} \BibitemShut {NoStop}%
\bibitem [{\citenamefont {Arun}\ \emph {et~al.}(2009)\citenamefont {Arun}, \citenamefont {Blanchet}, \citenamefont {Iyer},\ and\ \citenamefont {Sinha}}]{Arun:2009mc}%
  \BibitemOpen
  \bibfield  {author} {\bibinfo {author} {\bibfnamefont {K.~G.}\ \bibnamefont {Arun}}, \bibinfo {author} {\bibfnamefont {L.}~\bibnamefont {Blanchet}}, \bibinfo {author} {\bibfnamefont {B.~R.}\ \bibnamefont {Iyer}},\ and\ \bibinfo {author} {\bibfnamefont {S.}~\bibnamefont {Sinha}},\ }\bibfield  {title} {\bibinfo {title} {{Third post-Newtonian angular momentum flux and the secular evolution of orbital elements for inspiralling compact binaries in quasi-elliptical orbits}},\ }\href {https://doi.org/10.1103/PhysRevD.80.124018} {\bibfield  {journal} {\bibinfo  {journal} {Phys. Rev. D}\ }\textbf {\bibinfo {volume} {80}},\ \bibinfo {pages} {124018} (\bibinfo {year} {2009})},\ \Eprint {https://arxiv.org/abs/0908.3854} {arXiv:0908.3854 [gr-qc]} \BibitemShut {NoStop}%
\bibitem [{\citenamefont {Arun}\ \emph {et~al.}(2008{\natexlab{a}})\citenamefont {Arun}, \citenamefont {Blanchet}, \citenamefont {Iyer},\ and\ \citenamefont {Qusailah}}]{Arun:2007rg}%
  \BibitemOpen
  \bibfield  {author} {\bibinfo {author} {\bibfnamefont {K.~G.}\ \bibnamefont {Arun}}, \bibinfo {author} {\bibfnamefont {L.}~\bibnamefont {Blanchet}}, \bibinfo {author} {\bibfnamefont {B.~R.}\ \bibnamefont {Iyer}},\ and\ \bibinfo {author} {\bibfnamefont {M.~S.~S.}\ \bibnamefont {Qusailah}},\ }\bibfield  {title} {\bibinfo {title} {{Tail effects in the 3PN gravitational wave energy flux of compact binaries in quasi-elliptical orbits}},\ }\href {https://doi.org/10.1103/PhysRevD.77.064034} {\bibfield  {journal} {\bibinfo  {journal} {Phys. Rev. D}\ }\textbf {\bibinfo {volume} {77}},\ \bibinfo {pages} {064034} (\bibinfo {year} {2008}{\natexlab{a}})},\ \Eprint {https://arxiv.org/abs/0711.0250} {arXiv:0711.0250 [gr-qc]} \BibitemShut {NoStop}%
\bibitem [{\citenamefont {Arun}\ \emph {et~al.}(2008{\natexlab{b}})\citenamefont {Arun}, \citenamefont {Blanchet}, \citenamefont {Iyer},\ and\ \citenamefont {Qusailah}}]{Arun:2007sg}%
  \BibitemOpen
  \bibfield  {author} {\bibinfo {author} {\bibfnamefont {K.~G.}\ \bibnamefont {Arun}}, \bibinfo {author} {\bibfnamefont {L.}~\bibnamefont {Blanchet}}, \bibinfo {author} {\bibfnamefont {B.~R.}\ \bibnamefont {Iyer}},\ and\ \bibinfo {author} {\bibfnamefont {M.~S.~S.}\ \bibnamefont {Qusailah}},\ }\bibfield  {title} {\bibinfo {title} {{Inspiralling compact binaries in quasi-elliptical orbits: The Complete 3PN energy flux}},\ }\href {https://doi.org/10.1103/PhysRevD.77.064035} {\bibfield  {journal} {\bibinfo  {journal} {Phys. Rev. D}\ }\textbf {\bibinfo {volume} {77}},\ \bibinfo {pages} {064035} (\bibinfo {year} {2008}{\natexlab{b}})},\ \Eprint {https://arxiv.org/abs/0711.0302} {arXiv:0711.0302 [gr-qc]} \BibitemShut {NoStop}%
\bibitem [{\citenamefont {Mishra}\ \emph {et~al.}(2015)\citenamefont {Mishra}, \citenamefont {Arun},\ and\ \citenamefont {Iyer}}]{Mishra:2015bqa}%
  \BibitemOpen
  \bibfield  {author} {\bibinfo {author} {\bibfnamefont {C.~K.}\ \bibnamefont {Mishra}}, \bibinfo {author} {\bibfnamefont {K.~G.}\ \bibnamefont {Arun}},\ and\ \bibinfo {author} {\bibfnamefont {B.~R.}\ \bibnamefont {Iyer}},\ }\bibfield  {title} {\bibinfo {title} {{Third post-Newtonian gravitational waveforms for compact binary systems in general orbits: Instantaneous terms}},\ }\href {https://doi.org/10.1103/PhysRevD.91.084040} {\bibfield  {journal} {\bibinfo  {journal} {Phys. Rev. D}\ }\textbf {\bibinfo {volume} {91}},\ \bibinfo {pages} {084040} (\bibinfo {year} {2015})},\ \Eprint {https://arxiv.org/abs/1501.07096} {arXiv:1501.07096 [gr-qc]} \BibitemShut {NoStop}%
\bibitem [{\citenamefont {Boetzel}\ \emph {et~al.}(2019)\citenamefont {Boetzel}, \citenamefont {Mishra}, \citenamefont {Faye}, \citenamefont {Gopakumar},\ and\ \citenamefont {Iyer}}]{Boetzel:2019nfw}%
  \BibitemOpen
  \bibfield  {author} {\bibinfo {author} {\bibfnamefont {Y.}~\bibnamefont {Boetzel}}, \bibinfo {author} {\bibfnamefont {C.~K.}\ \bibnamefont {Mishra}}, \bibinfo {author} {\bibfnamefont {G.}~\bibnamefont {Faye}}, \bibinfo {author} {\bibfnamefont {A.}~\bibnamefont {Gopakumar}},\ and\ \bibinfo {author} {\bibfnamefont {B.~R.}\ \bibnamefont {Iyer}},\ }\bibfield  {title} {\bibinfo {title} {{Gravitational-wave amplitudes for compact binaries in eccentric orbits at the third post-Newtonian order: Tail contributions and postadiabatic corrections}},\ }\href {https://doi.org/10.1103/PhysRevD.100.044018} {\bibfield  {journal} {\bibinfo  {journal} {Phys. Rev. D}\ }\textbf {\bibinfo {volume} {100}},\ \bibinfo {pages} {044018} (\bibinfo {year} {2019})},\ \Eprint {https://arxiv.org/abs/1904.11814} {arXiv:1904.11814 [gr-qc]} \BibitemShut {NoStop}%
\bibitem [{\citenamefont {Ebersold}\ \emph {et~al.}(2019)\citenamefont {Ebersold}, \citenamefont {Boetzel}, \citenamefont {Faye}, \citenamefont {Mishra}, \citenamefont {Iyer},\ and\ \citenamefont {Jetzer}}]{Ebersold:2019kdc}%
  \BibitemOpen
  \bibfield  {author} {\bibinfo {author} {\bibfnamefont {M.}~\bibnamefont {Ebersold}}, \bibinfo {author} {\bibfnamefont {Y.}~\bibnamefont {Boetzel}}, \bibinfo {author} {\bibfnamefont {G.}~\bibnamefont {Faye}}, \bibinfo {author} {\bibfnamefont {C.~K.}\ \bibnamefont {Mishra}}, \bibinfo {author} {\bibfnamefont {B.~R.}\ \bibnamefont {Iyer}},\ and\ \bibinfo {author} {\bibfnamefont {P.}~\bibnamefont {Jetzer}},\ }\bibfield  {title} {\bibinfo {title} {{Gravitational-wave amplitudes for compact binaries in eccentric orbits at the third post-Newtonian order: Memory contributions}},\ }\href {https://doi.org/10.1103/PhysRevD.100.084043} {\bibfield  {journal} {\bibinfo  {journal} {Phys. Rev. D}\ }\textbf {\bibinfo {volume} {100}},\ \bibinfo {pages} {084043} (\bibinfo {year} {2019})},\ \Eprint {https://arxiv.org/abs/1906.06263} {arXiv:1906.06263 [gr-qc]} \BibitemShut {NoStop}%
\end{thebibliography}%

\end{document}